\documentclass[aps,showpacs,twocolumn]{revtex4}
\usepackage{amsmath}
\usepackage{amsfonts}
\usepackage{amssymb}
\usepackage{graphicx}

\usepackage{indentfirst}

\begin{document}


\title{Dynamics of recollisions for the double ionization of atoms in intense laser fields}

\author{F. Mauger$^1$, C. Chandre$^1$, T. Uzer$^2$}
\affiliation{$^1$ Centre de Physique Th\'eorique, CNRS -- Aix-Marseille Universit\'e, Campus de Luminy, case 907, F-13288 Marseille cedex 09, France \\  $^2$ School of Physics, Georgia Institute of Technology, Atlanta, GA 30332-0430, USA}
\date{\today}

\begin{abstract}
We investigate the dynamics of electron-electron recollisions in the double ionization of atoms in strong laser fields. The statistics of recollisions can be reformulated in terms of an area-preserving map from the observation that the outer electron is driven by the laser field to kick the remaining core electrons periodically. The phase portraits of this map reveals the dynamics of these recollisions in terms of their probability and efficiency.  
\end{abstract}
\maketitle



\section{Introduction}

When subjected to short and intense laser pulses, the helium atom (or other atoms or molecules with two active electrons) may undergo double ionization~\cite{Beck08}. The conventional route for double ionization is a sequential mechanism in which the field ionizes one electron after the other in an uncorrelated way. This process, called sequential double ionization (SDI)~\cite{Beck08}, allows simple theoretical predictions of double ionization yields~: The double ionization probability is given by the product of the single ionization probability with the probability of ionization of the remaining ion. However, experiments carried out using intense linearly polarized laser fields ~\cite{Fitt92,Kond93,Walk94,Laro98,Webe00,Corn00,Guo01,DeWi01,Ruda04} have shown that in the range of intensity between $10^{14}$ and $10^{15} \ {\rm W} \cdot {\rm cm}^{-2}$, double ionization yields depart significantly from the sequential predictions by several orders of magnitude. This observation has led to the identification of an alternative route to double ionization, called non-sequential double ionization (NSDI)~\cite{Beck08}, in which the correlation between the two electrons cannot be neglected. Today NSDI is regarded as one of the most dramatic manifestations of electron-electron correlation in nature. 

Various mechanisms have been proposed to explain this surprise~\cite{Fitt92, Cork93, Scha93, Walk94, Beck96, Kopo00, Lein00, Sach01, Fu01, Panf01, Barn03, Colg04, Ho05_1, Ho05_2, Ruiz05, Horn07, Prau07, Feis08}. When confronted with near-infrared experiments~\cite{Brya06,Webe00}, 
the recollision scenario~\cite{Cork93,Scha93} seems in best accord with  observations~\cite{Fitt92,Kond93,Walk94,Laro98,Webe00,Corn00,Guo01,DeWi01,Ruda04,Ivan05} and is validated by quantum~\cite{Lein00, Wats97}, semi-classical~\cite{Fu01, Chen03, Brab96} and classical simulations~\cite{Panf03, Fu01, Sach01, Ho05_1, Ho05_2, Panf02, Liu07, Maug09}~: According to this scenario, a pre-ionized electron (referred as the ``outer'' electron~\cite{Maug09}), after picking up energy from the laser field, is hurled back at the parent ion by the laser and collides (thereby exchanging a significant amount of energy) with the remaining electron (referred as the ``inner'' electron~\cite{Maug09}) trapped close to the nucleus. In general the inner electron experiences multiple recollisions and can eventually ionize, leading to double ionization if the outer electron remains ionized itself. Recollision has become ``the keystone of strong-field physics''~\cite{Beck08} in the understanding of the electronic dynamics and light source design~\cite{Brab00}.

Even though the recollision mechanism is well settled in its broad outlines, some issues persist about the nature of collisions involved since not every recollision leads to double ionization (or does so right away). The efficiency of these collisions in transferring ionization energy during recollision have a direct bearing on the double ionization probabilities.

In this manuscript, we investigate the dynamics of the recollisions which lead to double ionization, in particular the energy exchange between the two electrons during successive recollisions. A number of NSDI features, obtained using classical models~\cite{Panf03, Fu01, Sach01, Ho05_1, Ho05_2, Panf02, Liu07, Maug09}, are in very good agreement with results from quantum mechanical simulations and from experiments~\cite{Beck08}. This agreement is ascribed to the prominent role of electron-electron correlation~\cite{Beck08,Ho05_1, Ho05_2}. In addition, classical mechanical models have been used to a better understanding of the mechanisms because of their favorable scaling with system size. 

\begin{figure}[htb]
	\centering
		\includegraphics[width = \linewidth]{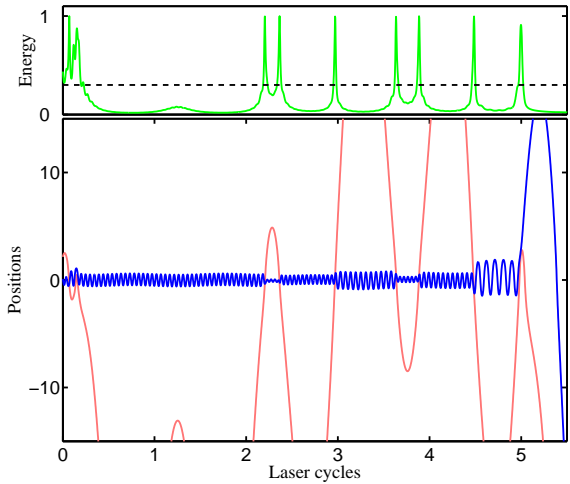}
	\caption{\label{fig:NSDI}
	Lower panel~: Positions of the two electrons of a typical double ionizing trajectory of Hamiltonian~(\ref{eq:Hamiltonian}) as a function of time. The parameters of the laser field are $I = 2 \times 10^{14} \ {\rm W} \cdot {\rm cm}^{-2}$ and $\omega = 0.0584$ a.u.
	Upper panel~: Interaction energy between the two electrons (green curve) defined as the soft Coulomb potential between the electrons. The dashed black line represents the threshold for recollision detection used in the analysis of recollisions.}
\end{figure}

In what follows we consider the following Hamiltonian system describing, in the dipole approximation, a one--dimensional He atom using soft Coulomb potentials~\cite{Java88, Ho05_1, Ho05_2, Panf02} driven by a linearly polarized laser field of amplitude $E_0$ and frequency $\omega$~:
\begin{eqnarray} \label{eq:Hamiltonian}
   H \left( x, y, p_{x}, p_{y}, t \right) = \frac{p_{x}^{2}}{2} + \frac{p_{y}^{2}}{2} 
      + \frac{1}{\sqrt{ \left( x-y \right)^{2} + 1}} \nonumber \\
      -\frac{2}{\sqrt{x^{2}+1}} - \frac{2}{\sqrt{y^{2}+1}} + (x+y) E_{0} \sin  \omega t,
\end{eqnarray}
where~$x$ and~$y$ denote the position of each electron, and~$p_{x}$ and~$p_{y}$ their~(canonically) conjugate momenta. The duration of the pulse, i.e., the duration of the time-integration of trajectories, is 8 laser cycles. An analysis of typical trajectories of Hamiltonian~(\ref{eq:Hamiltonian}) shows that the pre-ionized electron comes back to the inner region and exchanges energy with the inner electron several times which can be seen as repeated kicks delivered by the outer electron on the inner one (as seen in Fig.~\ref{fig:NSDI}). The key feature for double ionization is the energy exchanged during each kick. Viewing the recollision process as a periodic sequence of kicks suggests the use an area-preserving map which is constructed from periodically kicked dynamics, and widely used in various physical contexts in physics~\cite{chaosbook}. The most prominent examples in atomic physics are the maps developed to model ionization of Rydberg atoms driven by microwave fields~\cite{Casa88, ChaosAtomPhys}. 

Discrete-time models for continuous-time periodic processes enjoy great popularity in physics. A paradigmatic example is the standard map~\cite{chaosbook}, the simplicity of which makes it an effective toy model for the study of chaotic properties in Hamiltonian systems. In addition, some rigorous properties can be derived, e.g., the transition from regular to chaotic behavior, the existence of elliptic periodic orbits, the persistence of rotational invariant tori, etc. Whether it is for integrating numerically such continuous-time processes or modeling physical phenomena, the main advantages of maps are that they can be integrated more easily than continuous flows and that their properties appear with more clarity due to their reduced-dimensional phase space (as exemplified by Poincar\'e sections of continuous flows). Consequently they allow a deeper understanding of the dynamics and the underlying phenomena. 

Even though the nature of the kicks in the double ionization mechanism differs from what is observed for Rydberg atoms~\cite{Casa88, ChaosAtomPhys}, here we construct a map which is similar to the standard map. Using apt action-angle variables for the inner electron dynamics, we construct the following map~\cite{Maug10_1} for the recollision dynamics~:
\begin{equation} \label{eq:Mapping}
   \begin{array}{ccl}
      A_{n+1}       & = & A_{n}/(1- \varepsilon \sin \varphi_{n}), \\
      \varphi_{n+1} & = & \varphi_{n} + T\sqrt{2} \exp(a A_{n+1})+\varepsilon\cos \varphi_n,
   \end{array}
\end{equation}
where $A_{n}$ and $\varphi_{n}$ are, respectively, the action and angle variables associated with the inner electron right before the $n^{th}$ recollision. The constant $a$ depends on the chosen atom, e.g., for He, $a=-9\sqrt{2}/16$. The parameter $\varepsilon$ is the strength of the kick and will be related to the exchanges of energy at the recollision. The period of the kicks is denoted by $T$. The analysis of the phase portraits of this map shed some new light on the recollision dynamics.

In Sec.~\ref{sec:recol}, we construct the map given by Eq.~(\ref{eq:Mapping}) from the analysis of recollisions experienced by the trajectories of Hamiltonian~(\ref{eq:Hamiltonian}). In Sec.~\ref{sec:num}, we analyze numerically this map in order to infer some properties of the recollision dynamics and on the nonsequential route to double ionization. 

In what follows, instead of SDI and NSDI, we will use the more general UDI (uncorrelated double ionization) and CDI (correlated double ionization)~\cite{Maug10_1}. This terminology comes from the observation that a recollision may put the inner electron into an almost-bound state which then takes a significant time (sometimes more than one laser cycle) to ionize. With the previous definition, these so-called ''recollision excitation with subsequent ionization''~\cite{Feue01, Rude04} events -- by no means rare -- would be labeled as SDI (because of the large time delay between the ionization of the two electrons) whereas they clearly correspond to a correlated process in the same way as NSDI. Here we consider CDI, where at least one recollision is needed for double ionization, and UDI where no recollision is needed.

Some of these results were anounced in a recent Letter~\cite{Maug10_1}.


\section{Discrete-time model for recollisions}
\label{sec:recol}

Without the laser field~($E_{0}=0$), typical trajectories associated with Hamiltonian~(\ref{eq:Hamiltonian}) are composed of an electron close to the nucleus (the ``inner'' electron) and one electron further away (the ``outer'' electron)~\cite{Maug09}. This observation follows from the existence of four weakly hyperbolic periodic orbits which organize the chaotic motion. When the laser is turned on, the outer electron is quickly ionized while the inner one experiences a competition between the laser excitation and the Coulomb interaction with the nucleus. For nonsequential double ionizations, the inner electron is trapped in a bound region about the nucleus and the only way to free itself is through a recollision with the outer electron when it returns to the core. We give an example of such a trajectory in Fig.~\ref{fig:NSDI}. We note the fast first ionization of one electron while the other one remains trapped close to the nucleus. Because of the laser oscillations, the pre-ionized electron is hurled back at the core and recollides repeatedly with the other one. After about 5 laser cycles, a final recollision manages to free both electrons and leads to a correlated double ionization.

In a nutshell, the mechanism is the following one~: When the outer electron returns to the core, the soft Coulomb interaction between the two electrons is no longer negligible (particularly for the inner electron). It results in a rearrangement for this electron for which the action is modified. Since the outer electron comes back to the core quickly, this interaction is approximated by a kick experienced by the inner electron~: When the outer electron returns to the core, it gives a kick in action to the inner one which jumps from one invariant torus to another (or to the unbound region, thereby ionizing). The action of the inner electron is constant between two recollisions.  

In this section, we construct a simplified model for the recollisions which comes down to the following kicked Hamiltonian~\cite{Maug10_1}:
\begin{equation} \label{eq:HamiltMap}
      H_{\rm m} \left( \varphi, A,t \right)  = H_0(A) + 
         \varepsilon A \cos \varphi \ \sum_{n=1}^{N} {\delta \left( t - nT \right)},
\end{equation}
where~$H_0(A)$ is the integrable part of the Hamiltonian of the inner electron, and $T$ is the delay between two recollisions, and $\varepsilon$ represents the strength of the kick and only depends on the intensity of the laser. Each part of Hamiltonian (\ref{eq:HamiltMap}) is designed from theoretical models and supported by statistical analysis of the recollisions.

We construct an area-preserving map (From the kicked Hamiltonian given by Eq.~(\ref{eq:HamiltMap})). This is done in the standard way~\cite{chaosbook} by defining $\varphi_{n}$ and~$A_{n}$ as the angle and action of a trajectory of Hamiltonian~(\ref{eq:HamiltMap}) at time~$\left(nT\right)^{-}$~(right before the $n^{th}$ kick). By integrating the trajectories between two kicks, i.e.,\ from $t=(nT)^-$ to $t=[(n+1)T]^-$, we approximate the dynamics of Hamiltonian~(\ref{eq:HamiltMap}) by the two-dimensional symplectic map~(\ref{eq:Mapping}).


\subsection{Integrable component~: $H_{0} \left( A \right)$}
\label{sec:IIA}

The effective Hamiltonian for the inner electron is given by~\cite{Maug09}~:
\begin{equation} \label{eq:InnerHamilt}
   H_{\rm in} \left( y , p_{y},t \right) = \frac{p_{y}^{2}}{2} - \frac{2}{\sqrt{y^{2}+1}} 
      + y E_{0} \sin \omega t,
\end{equation}
which is obtained from Hamiltonian~(\ref{eq:Hamiltonian}) by neglecting the interaction with the other electron. A quick inspection at its phase space shows that there are two main regions which result from the competition between the laser field and the Coulomb interaction~: a bound region close to the nucleus where the electron remains bounded (on invariant tori) at all times, and an unbound region where the electron leaves the nucleus quickly and ionizes~\cite{Maug09}.   
The distance of the inner electron around the nucleus is best expressed in terms of its energy  
\begin{equation} \label{eq:H0py}
   H_0 \left(y, p_{y} \right)=\frac{p_{y}^{2}}{2} - \frac{2}{\sqrt{y^{2}+1}}. 
\end{equation}
The smaller this energy, the closer is the electron to the nucleus. In the absence of the field, $H_0$ gives a natural criterion for ionization, based on energy conservation for autonomous Hamiltonian systems. If the energy is smaller than zero, then the Coulomb interaction with the nucleus is strong enough to maintain the electron at a finite distance for all times. On the contrary, if the energy is positive, then the electron will escape to infinity.

\begin{figure}[htb]
	\centering
		\includegraphics[width = \linewidth]{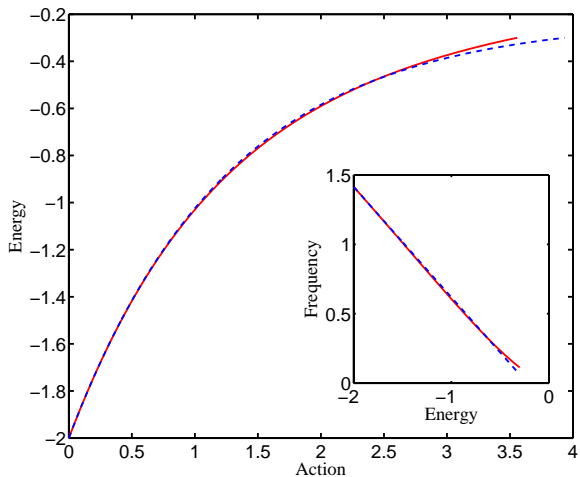}
	\caption{\label{fig:H0} Numerical evaluation of the energy given by Eq.~(\ref{eq:H0py}) as a function of the action $A$ (continuous red curves). The inset displays the frequency of the inner electron versus its energy. For comparison, in both  panels, we give the values predicted by the approximate model~(\ref{eq:H0}) (dashed blue curves).}
\end{figure}

In the neighborhood of the nucleus, the motion is harmonic with a frequency of $\sqrt{2}$, and moving away from the nucleus, the frequency decreases. We observe numerically that the frequency $\nu$ associated with $H_0$ depends approximately linearly on the energy in the whole bound region (see Fig.~\ref{fig:H0}, inner panel)~:
$$
   \nu \left( \mathcal{E} \right) = a \left( \mathcal{E} + 2 \right) +  \sqrt{2},
$$
where $\mathcal{E}$ is the energy of the inner electron. We perform a change of coordinates into action-angle in the system described by Hamiltonian~(\ref{eq:H0py}). From the equation $\nu(H_0)=\partial H_0/\partial A$, we obtain an expression for $H_0$~:
\begin{equation} \label{eq:H0}
   H_0(A)=-2+\sqrt{2}\frac{{\rm e}^{aA}-1}{a}.
\end{equation}

The parameter $a$ can be computed from a series expansion of the action associated with the inner electron around the bottom of the well. The equation for the action is
\begin{equation} \label{eq:Action_def}
   A = \frac{1}{2 \pi} \oint{p_{y} \ dy} = \frac{1}{\pi} \int_{-y_{m}}^{y_m}{\sqrt{\frac{4}{\sqrt{y^{2}+1}} + 2 \mathcal{E}} dy},
\end{equation}
where $y_{m} = \sqrt{4/\mathcal{E}^2 - 1}$ is the maximum position the electron can experience when it has energy $\mathcal{E}$. We define the energy as $\mathcal{E} = -2\sqrt{1 - \xi^{2}}$, where $\xi$ is a small (positive) parameter. Then, considering a series expansion in $\xi$, we end up with
$$
   a = -\frac{9 \sqrt{2}}{16}.
$$
Numerically, $a \approx -0.8$~\cite{Maug10_1}. In Fig.~\ref{fig:H0}, we compare the value for the action given by Eq.~(\ref{eq:H0}) with a numerical evaluation of the integral~(\ref{eq:Action_def}).


\subsection{The kicks}


\subsubsection{Time delay between recollisions and number of recollisions}

The effective Hamiltonian for the outer electron is
$$
   H_{{\rm out}} \left( x, p_{x}, t \right) = \frac{p_{x}^{2}}{2} + x E_{0} \sin \omega t,
$$
which is obtained from Hamiltonian~(\ref{eq:Hamiltonian}) by neglecting the interaction with the other electron and with the nucleus. 
Trajectories associated with Hamiltonian $H_{{\rm out}}$ are composed of a linear escape modulated by a sine function with the same period as the laser. It means that the outer electron typically experiences two returns to the core per laser cycle.

\begin{figure}[htb]
	\centering
		\includegraphics[width = \linewidth]{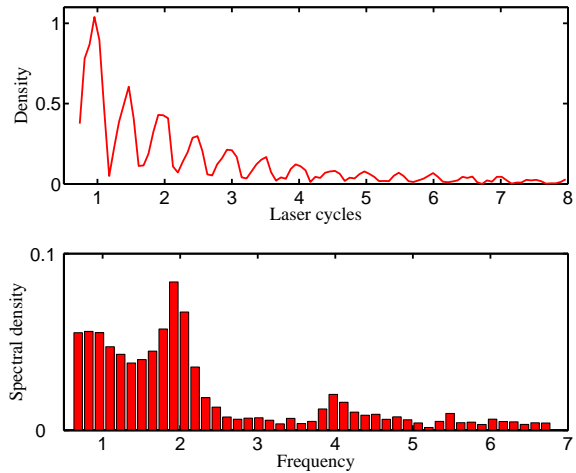}
	\caption{\label{fig:Recollison_times}
	Densities of recollision times (upper panel) and its Fourier decomposition (lower panel) obtained from a statistical analysis of a large assembly of trajectories for an intensity $I = 2 \times 10^{14} \ {\rm W} \cdot {\rm cm}^{-2}$ and laser frequency $\omega = 0.0584$ a.u..}
\end{figure}

\begin{figure}[htb]
	\centering
		\includegraphics[width = \linewidth]{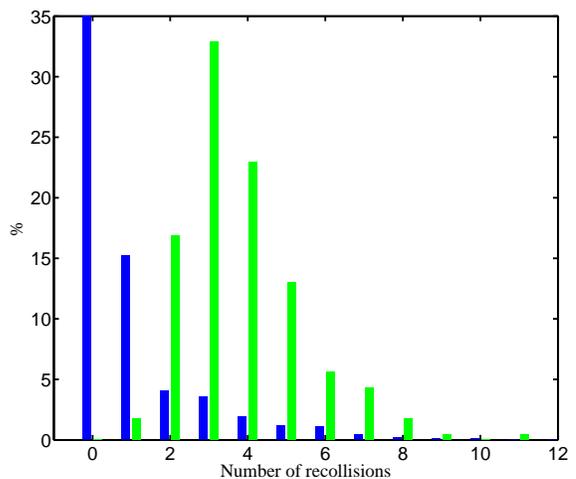}
	\caption{\label{fig:NbRec_density}
	Distribution of the numbers of recollisions for a laser intensity $I = 2 \times 10^{14} \ {\rm W}\cdot{\rm cm}^{-2}$ and frequency $\omega = 0.0584$ a.u.. Dark blue bars correspond to data collected from typical trajectories (including non-ionizing ones) and light green bars correspond to data associated with double ionizing trajectories. For a better layout, the middle part of the statistics has been cut for the zero recollision component which goes up to 70\%.}
\end{figure}

We have collected statistical data from 16000 trajectories associated with Hamiltonian~(\ref{eq:Hamiltonian}) at recollision times, starting with a microcanonical initial distribution over the ground state energy surface $H(t=0)=-2.24$ a.u.~\cite{Haan94, Ho05_1, Maug09}. Numerically looking at the peaks in the energy of interaction between the electrons (defined as the soft Coulomb potential between the electrons) reveals times of recollision (see upper panel of Fig.~\ref{fig:NSDI}). From there on, one can collect and analyze characteristic data of the recollision process such as times when they take place, momentum of the outer electron, number of recollisions and exchanged energy (or action) during non-ionizing recollisions. One can define an energy for the inner electron, defined by $H_{0}$, as long as it is trapped in the bound region; it results in the impossibility to measure the amount of exchanged energy during recollisions leading to ionization. Consequently, ionizing recollisions are systematically discarded from the statistical analysis. In Fig.~\ref{fig:Recollison_times}, we give an example of the density of return times of the outer electron and its associated spectral decomposition.
It reveals that the main frequency for recollisions peaks around 2 per laser cycle which corresponds to a main period between two recollisions of half a laser cycle.  As a result, in the map we set the time delay between successive kicks to be equal to $T = \pi/\omega$.
Since the time duration of the laser pulse is 8 laser cycles, the inner electron experiences at most 15 recollisions. In Fig.~\ref{fig:NbRec_density}, we display the statistical distribution of the number of recollisions collected from the analysis of typical trajectories and typical double ionizing ones, for a fixed intensity. It shows that most of the trajectories do not undergo any recollision and typically between 2 to 4 recollisions are required to trigger correlated double ionization. The density depends weakly on the intensity in the intermediate range of intensity. It indicates that the map should not be iterated more than 12 times to reproduce accurately the dynamics of the recollisions experienced by the inner electron.

\subsubsection{Exchanged action during recollisions}

In Hamiltonian~(\ref{eq:HamiltMap}), the recollisions are modeled by a kick in action equal to $\varepsilon A \cos\varphi$ such that a kick might increase or decrease the action according to the respective phase between the two electrons. In addition, it is more difficult to kick the inner electron out if it is at the bottom of the well, so the kick strength is proportional to $A$. In this way, the action remains positive at all times since $A=0$ is invariant. The maximum strength of the kick depends strongly on $E_0$.

\begin{figure}[htb]
	\centering
		\includegraphics[width = \linewidth]{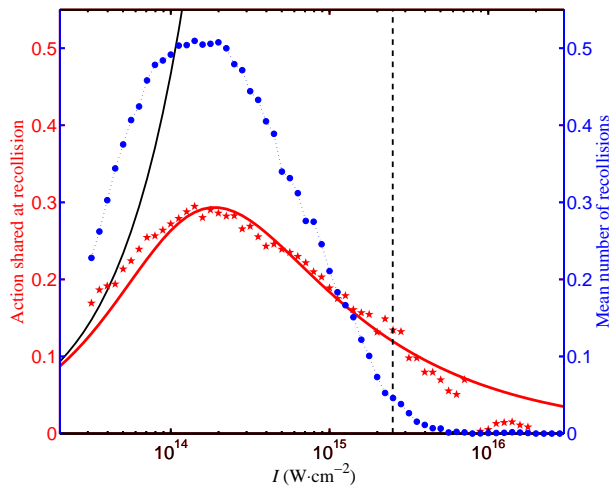}
	\caption{\label{fig:EnergySharing}
	Standard deviation~(red stars, left hand vertical scale) of shared action during recollisions as a function of the laser intensity $I$. An approximation of the standard deviation~(red continuous curve) is given by Eq.~(\ref{eq:ExchangedEnergyFit}). The thin black continuous line corresponds to the maximum recollision energy $\mathcal{E}_{\max}$ (expressed in terms of actions) and the vertical dotted line indicates the intensity after which we stop iterating the map. The dotted curve (blue line, right-hand vertical scale) shows the mean number of recollisions computed from all analyzed trajectories.}
\end{figure}

It is well-known that the maximum energy the outer electron can bring back to the core is equal to $\mathcal{E}_{\max}=\kappa U_p$ where $U_p=E_{0}^{2}/\left(4\omega^{2}\right)$ is the ponderomotive energy and $\kappa\approx 3.17$~\cite{Cork93, Scha93, Band05}. However, an inspection of the  energy exchanged during recollisions shows that this amount is significantly smaller (see Fig.~\ref{fig:EnergySharing}). Below, we analyze the recollisions in order to explain the trends observed in Fig.~\ref{fig:EnergySharing}.

\paragraph{Low intensity limit}

For an inner electron at the bottom of the well ($A=0$) which experiences a kick with energy $\kappa U_{p}$, its net change in action is thus equal to $\Delta A 
= \log \left( 1+ \kappa U_{p} a/\sqrt{2} \right)/a$, using Eq.~(\ref{eq:H0}). To leading order it gives
$$
   \Delta A \approx \frac{\kappa}{4 \sqrt{2} \omega^{2}} E_{0}^{2},
$$
for small $E_0$. 
It confirms the increase as $E_{0}^{2}$ observed through the analysis of recollisions when the amplitude of the laser is weak (see Fig.~\ref{fig:EnergySharing}).

\paragraph{High intensity limit}
\label{sec:LargeIntLimit}

\begin{figure}[htb]
	\centering
		\includegraphics[width = \linewidth]{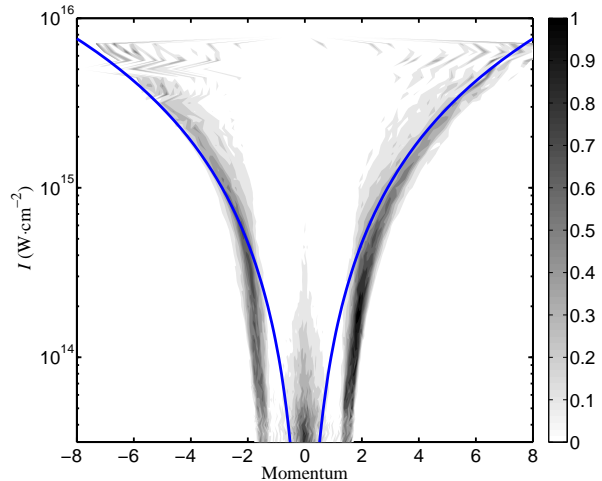}
	\caption{\label{fig:Outer_Electron_Momentum}
	Outer electron momentum distribution at recollision times as a function of the intensity of the field. The density is obtained from the analysis of a large assembly of trajectories of Hamiltonian~(\ref{eq:Hamiltonian}) for $\omega=0.0584$ a.u.. For comparison we also display momenta $p_{0} = \pm \gamma' \kappa' E_{0}$ where $\gamma' = 0.8$ (blue line).}
\end{figure}

As the intensity of the laser field increases up to $10^{14}\ {\rm W}\cdot{\rm cm}^{-2}$, the energy the outer electron brings back to the core (and potentially to the inner electron) varies approximately as $E_{0}^{2}$. Increasing the intensity further, it appears from the analysis of recollisions that the exchanged action decreases after a critical intensity of $2\times 10^{14}\ {\rm W}\cdot{\rm cm}^{-2}$. To model the process of recollision, we consider a Hamiltonian for the inner electron ($y$) when the outer one ($x$) comes back to the core as~:
\begin{eqnarray} 
  && H_{{\rm rec}} \left(y, p_{y}, t\right) = \frac{p_{y}^{2}}{2} - \frac{2}{\sqrt{y^{2}+1}} \nonumber \\
  && \quad + \frac{1}{\sqrt{ \left(y-x\left( t \right)\right)^{2} + 1 }} + y E_{0} \sin \omega t, \label{eq:RecollisionHamilt}
\end{eqnarray}
which is obtained from the reduced Hamiltonian~(\ref{eq:InnerHamilt}) of the inner electron by adding a passive soft Coulomb interaction with a quickly returning outer electron. 
Because of its large momentum, we consider that the outer electron is not affected by the interaction with the inner one (i.e.\ we impose the dynamics for the outer electron independently of the dynamics of the inner one) and the nucleus. From its position and momentum at recollision, respectively $y_{0}$ and $p_{0}$, its trajectory is almost a straight line~: $x \left( t \right) = y_{0} + p_{0} t$, where we have set the origin of times at the time of the recollision. 
For large intensities, the inner electron in the bound region is close to the nucleus, since the bound region becomes smaller, and in such a configuration the dynamics is well approximated by harmonic potentials. Finally, we assume the motion for both the inner and outer electrons to be much faster than the laser field variation. Thus it is assumed that locally $E_{0} \sin \omega t \approx  \eta E_{0}$. As a result, Hamiltonian $H_{{\rm rec}}$ given by Eq.~(\ref{eq:RecollisionHamilt}) is simplified through an expansion around the position $y=0$ as
\begin{equation} \label{eq:H_rec}
   H_{{\rm rec}} \left( y, p_{y}, t \right) = \frac{p_{y}^{2}}{2} + y^{2} - \frac{1}{2} \left( y - x \left( t\right) \right)^{2} + \eta E_{0} y.
\end{equation}
This model of recollision is valid as long as the two electrons are close enough to each other to allow an effective exchange of energy. We denote by $L$ the maximum distance between them to have an effective interaction. When the outer electron is further away, the term $\left( y - x \right)^{2}$ is canceled out. The resulting interval of time $\left( -\tau , \tau  \right)$ during which the two electrons are interacting with each other is equal to $\tau = L/p_{0}$. When the outer electron leaves the region of interaction, the effective model for the inner one becomes~:
$$
   H_{{\rm har}} = \frac{p_{y}^{2}}{2} + y^{2} + \eta E_{0} y.
$$ 
The resulting exchanged energy during the recollision is equal to $\Delta H = H_{{\rm har}} \left( y \left( \tau \right), p_y \left( \tau  \right) \right) - H_{{\rm har}} \left( y \left( -\tau  \right), p_y \left( -\tau  \right) \right)$. The positions $y \left( \pm \tau \right)$ and momenta $p_y \left( \pm \tau \right)$ are computed by (forward and backward) integration of the trajectory with initial conditions $y \left( t=0 \right) = y_{0}$ and $p_y \left( t= 0 \right) = \tilde{p}_{0}$ (the inner electron momentum at recollision) and whose dynamics is given by Hamiltonian~(\ref{eq:H_rec}). We expand these expressions up to order $O(\tau^4)$, and the leading term of $\Delta H$ is given by $2p_0\eta E_0\tau^3/3$. From the recollision  picture~\cite{Cork93,Band05} the maximum momentum the outer electron can have when it returns to the core is $p_{0}^{\max} = \kappa' E_{0}$, where $\kappa' = \sqrt{\kappa/(2 \omega^{2})}$. Statistical analysis of trajectories shows that for large intensities, the momentum distribution of the outer electron is centered around $\pm \gamma' p_{0}^{\max}$ where $\gamma' \approx 0.8$ (see Fig.~\ref{fig:Outer_Electron_Momentum}). As a result, we choose the outer electron's momentum proportional to the field amplitude~: $p_{0} = \gamma E_{0}$ (i.e.\ $\gamma = \gamma' \kappa'$) and the trajectory of the inner electron during recollision can be computed analytically. It allows one to consider a series expansion for the exchanged energy for the inner electron. From $\Delta H$, we recover the net variation in action at the recollision through Eq.~(\ref{eq:H0})~:
$$
   \Delta A = \frac{\Delta H}{\sqrt{2}} + \mathcal{O} \left( H^{2} \right).
$$
To leading order, the net exchange in action is equal to~:
$$
   \Delta A = \frac{\sqrt{2} \eta L^{3}}{3 \gamma^2} \frac{1}{E_{0}} + \mathcal{O} \left( \frac{1}{E_{0}^{2}} \right).
$$
This simple model of recollision explains the decrease as $1/E_{0}$ for the action exchange observed during the analysis of recollision, provided that the interaction length $L$ is independent of $E_0$. Moreover, $\Delta A$ varies as $\omega^2$, which means that we expect less exchanged energy at low frequency for high intensities. So the correlated double ionization probabilities are expected to be lower in the high intensity regime, and higher for low intensities as the laser frequency decreases.   

In summary, to combine the two trends of the mean shared action $\Delta A\left(E_{0}\right)$ (proportional to~$E_{0}^{2}$ at low intensities as given by ${\cal E}_{\max}$ and to~$1/E_{0}$ at higher ones) we fit it by~:
\begin{equation} \label{eq:ExchangedEnergyFit}
   \Delta A \left( E_{0} \right) = \frac{\alpha E_{0}^{2}}{1 + \beta E_{0}^{3}}.
\end{equation}
The parameter~$\alpha$ is equal to $\kappa/(4\sqrt{2}\omega^2)$. The parameter $\beta$ is given by $\beta=3\kappa^2\gamma'^{2}/(16\eta L^4\omega^2)$ and is obtained by a numerical fit so as to accurately reproduce the evolution of the mean exchanged action during recollisions (see continuous lines in Fig.~\ref{fig:EnergySharing}). For instance, for $\omega=0.0584$ a.u., the fitted value for $\beta$ is $\beta=5.1\times 10^3$. We consider for the kick strength in our map, $\varepsilon = \Delta A \left( E_{0} \right)$ as given by Eq.~(\ref{eq:ExchangedEnergyFit}).
 

\section{Numerical analysis of the map~(\ref{eq:Mapping})}
\label{sec:num}


\subsection{Inner electron distribution}

\begin{figure}[htb]
	\centering
		\includegraphics[width = \linewidth]{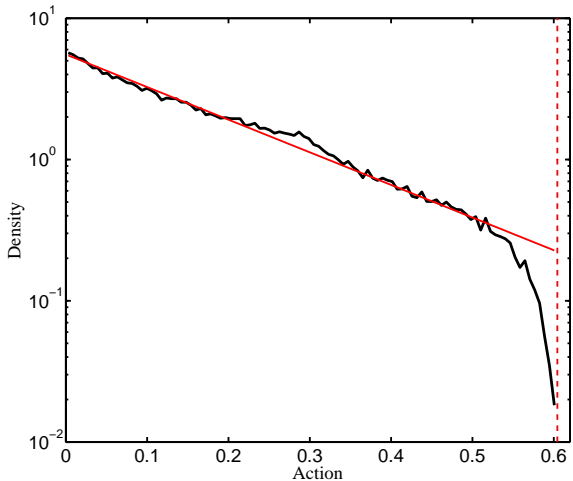}
	\caption{\label{fig:InitDistrib}
	Inner electron distribution function versus action. We compare the density obtained from a microcanonical set of $10^{5}$ initial conditions (large black curve) with the exponential law (\ref{eq:InnerDistribution}) (thin red curve). For comparison, we display the initial maximum action for the inner electron $A_{0}$ (dashed red line).}
\end{figure}

Starting from a microcanonical distribution on the ground state energy surface one can identify an inner and an outer electron~: the inner one is the electron with the smallest energy. We compute the distribution in action for the inner electron as defined in Sec.~\ref{sec:IIA}. An inspection of the shape of this density reveals that it has an exponential decrease on the range of accessible actions (see Fig.~\ref{fig:InitDistrib}). To model this distribution, we choose an exponential law with a truncated tail given by
\begin{equation} \label{eq:InnerDistribution}
   f_{_{\lambda, A_0}} \left( A \right) = \frac{\lambda}{1 - {\rm e}^{-\lambda A_0}} {\rm e}^{-\lambda A} \chi_{_{\left[0, A_0 \right]}} \left( A \right),
\end{equation}
where $A_0$ is the maximum allowed action for the inner electron and $\chi_{_{\Omega}} \left( x \right) = 1$ if $x \in \Omega$ and 0 otherwise. The two parameters $\lambda$ and $A_0$ are adapted so as to agree with the distribution deduced from the microcanonical set. The first one, $A_0$, is the maximum allowed action for the inner electron and it is computed from Hamiltonian~(\ref{eq:Hamiltonian}) for $E_0=0$. For the parameter $\lambda$, we consider a numerical fitting with a large assembly of initial conditions on the ground state energy surface. The fitting is done using the maximum likelihood method~\cite{Fish22} (the mean value method gives the same result). The numerical evaluation of the two parameters yields $A_0 = 0.61$ and $\lambda = 5.3$. In Fig.~\ref{fig:InitDistrib} we compare densities in action for the inner electron obtained from a microcanonical initial distribution and the exponential law given by Eq.~(\ref{eq:InnerDistribution}). Note the very good agreement between the two distributions for almost all allowed actions.

\subsection{Phase portrait of map~(\ref{eq:Mapping})}

\begin{figure}[htb]
	\centering
		\includegraphics[width = \linewidth]{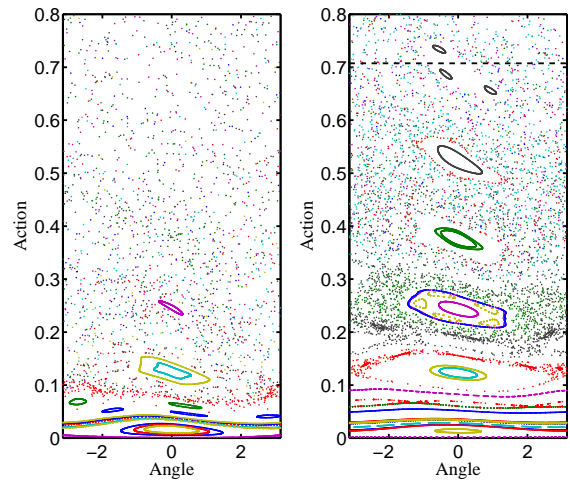}
	\caption{\label{fig:Mapping}
	Phase space portraits of some trajectories of the map~(\ref{eq:Mapping}) with $\omega=0.0584$ a.u. for low intensity (left panel) $I=2 \times 10^{14}\ \mbox{W}\cdot\mbox{cm}^{-2}$, and for high intensity (right panel) $I=2 \times 10^{15}\ \mbox{W}\cdot\mbox {cm}^{-2}$, represented by vertical dashed lines in Fig.~\ref{fig:knee}. In the right panel, we indicate the critical action~$A_{m}=0.71$, after which the inner electron ionizes, by a horizontal line~(whereas the critical action $A_{m}$ is $1.60$ in the left panel).}
\end{figure}

Now that the parameters of the map as well as the initial conditions are determined, we investigate numerically the dynamics given by map~(\ref{eq:Mapping}).
In Fig.~\ref{fig:Mapping}, we display two phase portraits for two laser intensities~: One  in the intermediate range of intensity ($I=2 \times 10^{14}\ \mbox{W}\cdot\mbox{cm}^{-2}$) shows a phase portrait which appears to be very chaotic, and one at high intensity ($I=2 \times 10^{15}\ \mbox{W}\cdot\mbox {cm}^{-2}$) shows a more regular phase portrait. In the chaotic region, the diffusion is much stronger in the intermediate range of intensities than for larger intensities (see Fig.~\ref{fig:Mapping}, left panel, where trajectories quickly escape from the core region, explaining why there are fewer points than in the right panel). Since the strength of the kicks decreases with the intensity at high intensity, the phase space becomes more regular. If the inner electron is inside an elliptic island (which occurs mainly at high intensities), it will not ionize regardless of the number of recollisions it undergoes. As the intensity increases, the recollisions become less effective and the map becomes integrable so fewer CDI events occur. Therefore there are two competing mechanisms for the vanishing of the CDI probability at high intensity~: the decrease of the size of the bound region, and the lack of efficiency of the recollisions due to the regularity of the dynamics of map~(\ref{eq:Mapping}). 

\subsection{Statistical analysis versus intensity}

Through map~(\ref{eq:Mapping}), we have derived a simple model for the dynamics of recollisions experienced by the inner electron initially in the bound region. From this model, we compute ionization probabilities of the inner electron from which we deduce the double ionization probability~: Double ionization occurs when the inner electron ionizes provided we assume that the outer electron remains ionized for all times. The picture of the bound and unbound regions for the effective Hamiltonian~(\ref{eq:InnerHamilt}) of the inner electron gives a natural criterion for ionization~: Once the inner electron has reached an action larger than the outermost invariant torus~(with action~$A_{m}$), it is driven away from the nucleus by the laser field. Therefore, all recollisions leading to an action larger than a critical value $A_m$ (which depends on $E_0$) subsequently lead to ionization of the inner electron. In angle-action variables, the unbound region becomes ~$\mathcal{D}\left(E_{0}\right)=\left\{\left(\varphi, A \right) \ \mbox{s.t.} \ A >A_{m}(E_0)\right\}$. We represent $A_m \left( E_{0} \right)$ as a function of the laser intensity in Fig.~\ref{fig:A_m}.

\begin{figure*}[htb]
	\centering
	  \includegraphics[width = .49\linewidth]{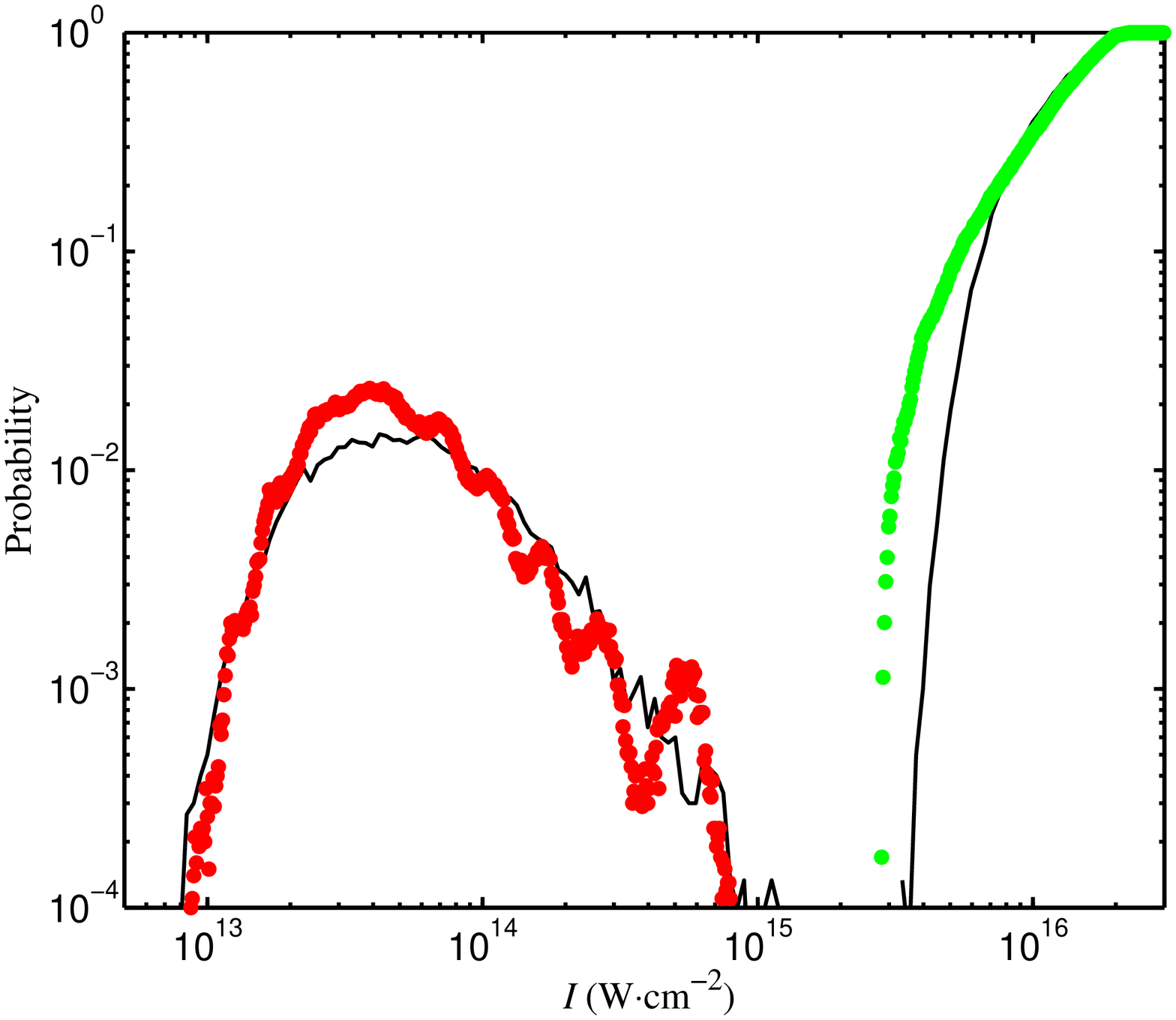}
		\includegraphics[width = .49\linewidth]{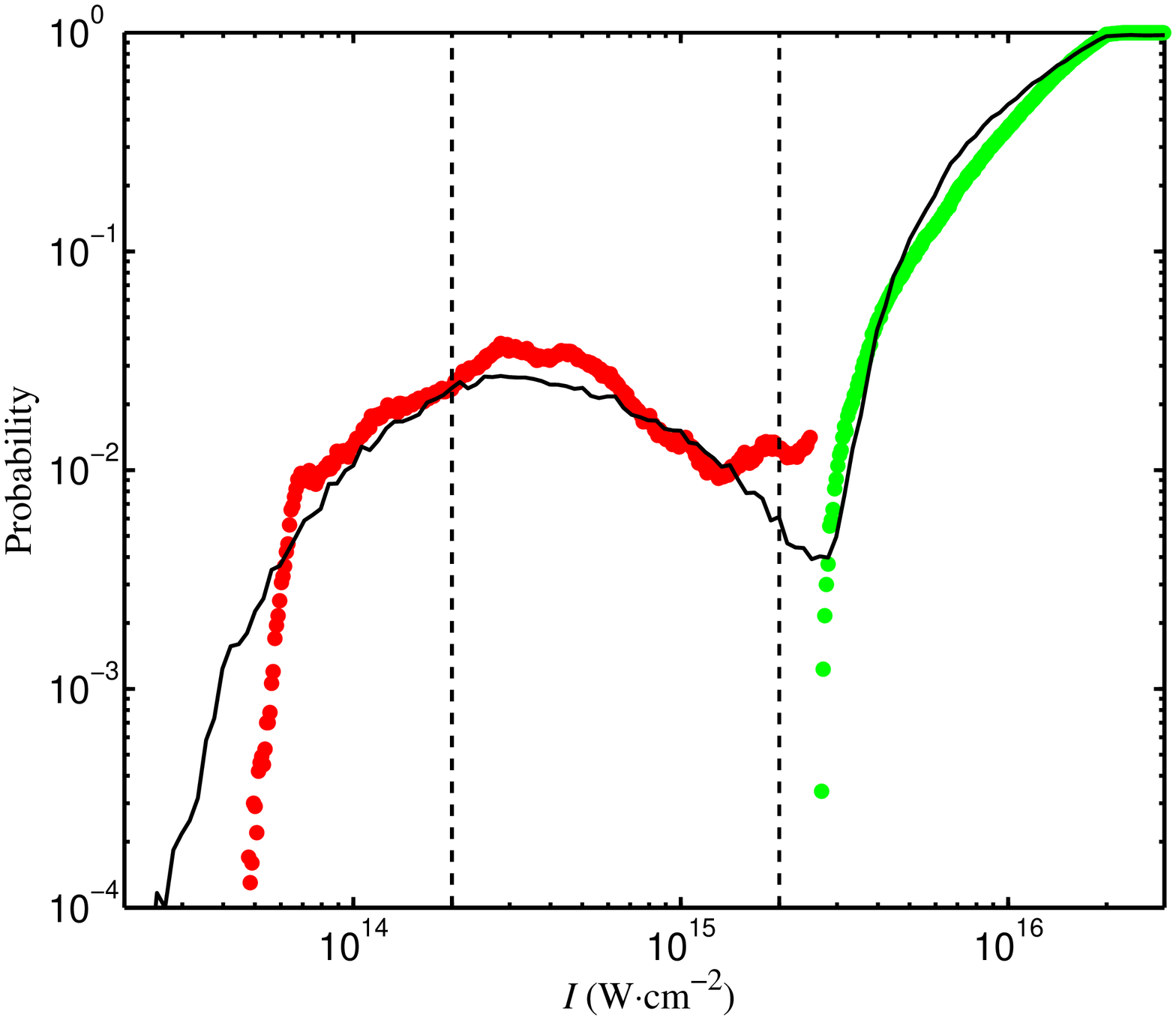}
	\caption{\label{fig:knee}
	Double ionization probability for Hamiltonian~(\ref{eq:Hamiltonian})~(continuous lines) and the map~(\ref{eq:Mapping})~(red dark circles) as a function of the laser intensity~$I$ for two laser frequencies (left panel $\omega = 0.0584$ a.u. and right panel $\omega = 0.02$ a.u.). Because of the low probability of recollision, we stop iterating the mapping after $I = 2.5 \times 10^{15} \ {\rm W} \cdot {\rm cm}^{-2}$ for $\omega = 0.0584$ a.u. and $I = 10^{15} \ {\rm W} \cdot {\rm cm}^{-2}$ for $\omega = 0.02$ a.u. For each frequency, we also indicate the expected UDI probability obtained as described in the text~(green light circles). Vertical dashed curves refer to intensities where phase portraits are displayed in Fig.~\ref{fig:Mapping}.}
\end{figure*}

\begin{figure}[htb]
	\centering
		\includegraphics[width = \linewidth]{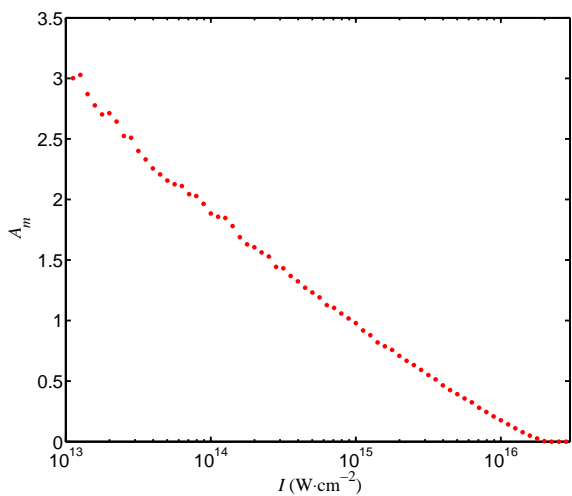}
	\caption{\label{fig:A_m}
	Action $A_{m} \left( E_{0} \right)$ of the last torus in the bound region of Hamiltonian~(\ref{eq:InnerHamilt}) as the laser intensity is varied for $\omega=0.0584$ a.u.}
\end{figure}

From an initial distribution in angle-action coordinates obtained from the microcanonical distribution~(\ref{eq:InnerDistribution}) of inner electrons in phase space, we iterate map~(\ref{eq:Mapping}) a fixed number of times for different intensities~(and thus different~$\varepsilon$). In what follows, we compare the results given by the map to the probability obtained using a direct integration of the trajectories of Hamiltonian~(\ref{eq:Hamiltonian}). As a function of the intensity of the field, these probability curves take the form of a ``knee''~\cite{Fitt92, Walk94, Webe00, Kond93, Corn00, Guo01, DeWi01, Ruda04, Beck96, Wats97, Lapp98, Panf03} which shows an enhancement of the double ionization probability in the intermediate range of intensities. In Fig.~\ref{fig:knee}, we display the double ionization probability as a function of the laser intensity.  We disregard recollisions for intensities larger than~$2.5 \times 10^{15} \ \mbox{W}\cdot\mbox {cm}^{-2}$. This adjustment is motivated by the weak probability of recollisions we have detected in the data analysis~(see Fig.~\ref{fig:EnergySharing}). A quick inspection of Fig.~\ref{fig:knee} reveals a bell-shaped curve for the resulting nonsequential component~\cite{Maug09, Maug10_1}. We notice that it qualitatively reproduces the trends observed in the double ionization yields observed using a statistical analysis of trajectories of Hamiltonian~(\ref{eq:Hamiltonian}) for two different values of the laser frequency. In particular, the asymmetry in the increase and decrease of the nonsequential component is worth noting.  

A rather intuitive mechanism to explain the decreasing part of this bell shape is the conversion from nonsequential trajectories into sequential ones when the laser field becomes stronger. However, in Fig.~\ref{fig:knee}, we notice a local decrease of the total yield (also observed with quantal computations~\cite{Lapp98}) which is larger by several orders than the increase of sequential process. We notice that this incompatibility is readily observed in Fig.~1 of Ref.~\cite{Lapp98}. As explained previously (see Sec.~\ref{sec:LargeIntLimit}), the decrease of the nonsequential component is mainly due to the decrease of recollision efficiency with the laser intensity.
At large intensity, the UDI probability is given by the proportion of the ground state energy surface where both electrons belong to $\mathcal{D} \left( E_{0} \right)$. 
In Fig.~\ref{fig:knee}, we display the UDI component predicted by this model (continuous blue curve) which is in good agreement with the double ionization probability obtained by integrating the full Hamiltonian~(\ref{eq:Hamiltonian}) in the high-intensity regime. 


\section{Conclusion}

In summary, we have studied key properties of the inelastic electron-electron recollisions through the analysis of recolliding trajectories. We connected our findings to a simplified model for the dynamics of the recollisions, which amounts to an area preserving map in the action-angle coordinates of the inner electron. The statistical analysis of the (discrete-time) trajectories of this model results in the hallmark ``knee'' shape for the probability of double ionization versus intensity. A proper decomposition into correlated and uncorrelated double ionization yields a  bellshape for the correlated process and a monotonic rise for the uncorrelated one. 

\acknowledgements
We acknowledge useful discussions with Adam Kamor. C.C. and F.M. acknowledge financial support from the PICS program of the CNRS. This work is partially funded by NSF.




\end{document}